# Data Provenance and Ethics in the Curricula of U.S. Health AI Education Programs

Raffaele Giancotti[1,¶*], Francesca Santucci[2¶], Amama Anjum[3,4], Monique de Villa[5], Rupsa Mitra[6], Soham Biren Katlariwala[7,8], Vishala Mishra[9], Leo Anthony Celi[1]

[1]MIT Critical Data, Massachusetts Institute of Technology, Cambridge (MA), USA

[2]The University of Texas at Austin, Austin (TX), USA

[3]Center of Artificial Intelligence, Prince Mohammad Bin Fahd University, Dhahran, Saudi Arabia

[4]Department of Computer Science, Government College University Faisalabad (GCUF), Faisalabad, Pakistan

[5]Department of Neurology, Neuroscape, University of California, San Francisco, San Francisco (CA), USA

[6]Centennial High School, Frisco (TX), USA

[7]Department of Health Science, University of the People, Pasadena (CA), USA

[8]Department of Computer Engineering, Gujarat Technological University, Ahmedabad, Gujarat, India

[9]Duke University School of Medicine, Durham (NC), USA

* Corresponding author: raf95@mit.edu (RG)

¶ These authors contributed equally to this work.

## Abstract

Failures of deployed clinical artificial intelligence (AI) systems have often originated in how their training data were generated. Whether the programs that train the health AI workforce teach practitioners to examine data provenance is unknown. We conducted a cross-sectional analysis of the publicly advertised curricula of health AI education programs at U.S. universities, identified through a structured multisource web search between April and July 2026. Two investigators coded coverage of data provenance (structural, dataset limitations, or data management only), ethics, and sustainability; collaborative structure; technical prerequisites; and the disciplinary background and gender of named instructors. The primary outcome was provenance coverage at the structural or dataset-limitations level. Of 43 candidate entries, 36 programs at 25 institutional units in 21 universities were included. Ethics was covered by 29 programs (80.6%; 95% confidence interval [CI], 65.0 to 90.2), data provenance by 8 (22.2%; 95% CI, 11.7 to 38.1), and structural provenance by 6 (16.7%; 95% CI, 7.9 to 31.9). None of the seven master's degree programs met the primary outcome, and no program addressed sustainability. Eighteen programs included team-based work, but only 5 assembled teams across clinical and technical backgrounds. Of 92 named instructors, 76.1% came from computational disciplines and 3.3% from the social sciences or humanities; among the 51 whose gender could be determined, 66.7% were men. U.S. health AI education programs widely taught ethics but rarely taught how clinical data are produced. None addressed the environmental cost of computation, and most instructors came from computational disciplines.

## Introduction

A sepsis prediction model embedded in the most widely used electronic health record in the United States was running at hundreds of hospitals before independent investigators found that it missed two thirds of sepsis cases and generated a heavy burden of false alarms [1]. An algorithm guiding care for millions of patients systematically underestimated the needs of Black patients because its designers chose health care cost as the training label, and unequal access to care made cost a poor proxy for illness [2]. Deep learning models can identify self-reported race from chest radiographs in ways their developers could not fully explain [3]. These problems originated upstream of the model, in the data and in the assumptions of those who selected them, and each was uncovered by investigators who had learned to ask where the data came from and what they measured.

Such habits are supposed to be formed during professional training. Universities now offer graduate degrees, certificates, executive programs, and online specializations to train the health AI workforce, yet what these programs teach has received far less scrutiny than what their graduates build. Reviews of AI education in the health professions report that only 58% of educational programs for medical trainees address the ethical, legal, and economic considerations of AI [4] and that existing guidance emphasizes evaluation and ethics over grounding in data science [5]. All 12 publications in one review recommended teaching AI ethics, but evidence on how to teach it remains scarce [6,7]. Ethics teaching has itself been reviewed [8], and a survey of 4596 students at 192 faculties in 48 countries found that 76.3% reported no AI courses [9]. Reviews of clinical datasets show that more than half of the datasets used for clinical AI originate in the United States or China [10]. Among publicly available clinical text datasets, 95.7% of those with a reported origin came from the Americas, Europe, or Asia and none from Africa or Oceania [11]. The two literatures rarely meet: curricular reviews do not treat data provenance as a distinct competency [7], and some fold dataset representativeness into ethics [8], so a claim that ethics is taught cannot be told apart from a claim that provenance is taught. What researchers know about how clinical data are produced has not made its way into what educators teach.

We examined 36 health AI education programs at U.S. universities and asked what they teach about data provenance, whether their ethics teaching connects to the origins of data, whether the environmental cost of computation appears at all, and whether programs recruiting across clinical, technical, and policy backgrounds structure collaboration among these learners. Health data are produced under conditions of unequal access and documentation [10], and practitioners enter a professional culture that devalues data work relative to model work [12]. Education succeeds only when its competencies match the needs of populations [13]. The programs we surveyed are among the most visible in the field and were not sampled at random. If provenance receives little attention even among them, the template from which formal AI education will scale does not include it. We present the findings as a baseline and plan to repeat the survey annually.

## Materials and methods

### *Study Design*

We conducted a cross-sectional analysis of the publicly advertised curricula of health AI education programs at U.S. universities. Such programs are advertised on institutional websites and are not indexed in bibliographic databases, so we used a structured multisource web search. Reporting follows the PRISMA extension for scoping reviews where applicable [14]. The review was not registered and no protocol was prepared in advance. The completed PRISMA-ScR checklist is provided in S2 File. The unit of analysis was the individual program or course offering. The study used only publicly available documents and involved no interaction with individuals, and instructor information is reported only in aggregate and de-identified form. We therefore did not consider it human subjects research and did not submit it for institutional review board review.

### *Search Strategy*

Candidates were identified between April and July 2026 from three sources: Google searches combining terms for AI, health care, education, and universities; two retrieval-augmented large language model assistants (Perplexity and Claude, accessed through their public web interfaces in June 2026, with model versions not recorded) prompted to enumerate U.S. university programs teaching AI in health care;

and hand searches of prior reviews of AI education. All queries and prompts are listed in Table S2 in S1 Appendix. Because model output is nondeterministic, the assistants were used only to generate candidates and never as evidence about any program. Every candidate was verified against its official institutional webpage, and all coded content derives from primary sources (program websites, syllabi, course catalogs, and faculty pages). Candidates were compiled into a screening register (S1 Data).

### *Eligibility and Selection*

Programs were eligible if they were offered by a U.S. university, medical school, or academic health center; had AI, machine learning, or data science applied to health as an explicit and substantial curricular focus; were open to enrollment at the time of the search as a degree, certificate, credit-bearing course, or structured professional program; and published at least a curriculum outline and a description of the intended audience. Offerings that had been discontinued or superseded by a currently taught version were excluded, and duplicate listings were merged. Programs run by distinct schools, campuses, or continuing-education arms of one university were counted as separate institutional units.

The search yielded 43 candidate entries. After 2 duplicates were merged, 5 of the 41 unique candidates were excluded (1 at a non-U.S. institution, 3 discontinued or superseded, and 1 general AI degree without health care content). The final sample comprised 36 programs at 25 institutional units in 21 universities (Fig 1 and Table S1). The flow diagram was adapted from the PRISMA 2020 template [15].

### *Data Extraction*

For each program, we extracted the level, format, duration, intended audience, admission prerequisites, curriculum structure, collaborative learning, and coverage of data provenance, ethics, and sustainability. Instructors were the individuals named as faculty, lecturers, or course directors on program pages; administrators, coordinators, and teaching assistants were excluded, and each person was counted once. Disciplinary background was assigned from the terminal degree and departmental appointment stated in the institutional biography, and gender from the pronouns used there; gender was recorded as unavailable when it could not be determined. De-identified instructor data are provided in Table S4 in S1 Appendix. When documentation was silent, variables were coded as not reported, distinct from an explicit no. Our question concerns what programs publicly commit to teaching, so programs coded as not reported were kept in the denominator.

### *Definitions*

We defined data provenance as the conditions under which health data are generated, including who is represented or absent and how clinical workflow and documentation shape the record, together with the limits these conditions place on what a model can learn. Coverage was coded from the content described, whether or not the word provenance appeared, at the highest level with explicit supporting evidence: structural provenance (the conditions of data generation, representation, or selection into the data); dataset limitations (technical consequences such as training-data bias, missingness, distribution shift, or limited generalizability, without their origins); or data management only (storage, standards, interoperability, database design, privacy, or quality assurance). The primary outcome was coverage at the structural or dataset-limitations level; a sensitivity analysis with narrower and broader definitions is reported in Table S3.

Ethics was coded as covered when any public documentation addressed ethics, responsible AI, fairness, bias, privacy, or regulatory compliance; depth was not graded, and content on the origins of data was coded under provenance only. Sustainability was coded as covered when documentation addressed the environmental, energy, or computational cost of AI systems. Collaborative structure was classified as team-based (listed team projects, capstones, or mentored group research), discussion only, none (explicitly stated), or not reported; team-based programs were classified as multidisciplinary when teams were explicitly assembled across clinical and technical backgrounds or paired with clinical mentors. Technical prerequisites were coded as required when admission required prior coursework in, or demonstrated proficiency with, programming, mathematics, statistics, or machine learning.

### *Coding and Statistical Analysis*

Two investigators independently coded all 36 programs from the same archived sources. Disagreements, most frequent for provenance and collaborative structure, were resolved by consensus with reference to the primary source; because pre-consensus assignments were not retained, chance-corrected agreement could not be computed. Results are reported as counts and percentages with 95% confidence intervals calculated with the Wilson score method. The sample comprises the programs we could identify and is not a probability sample.  No hypothesis tests were performed, and the widths of the confidence intervals have not been adjusted for multiplicity, so the intervals should not be used in place of hypothesis testing.

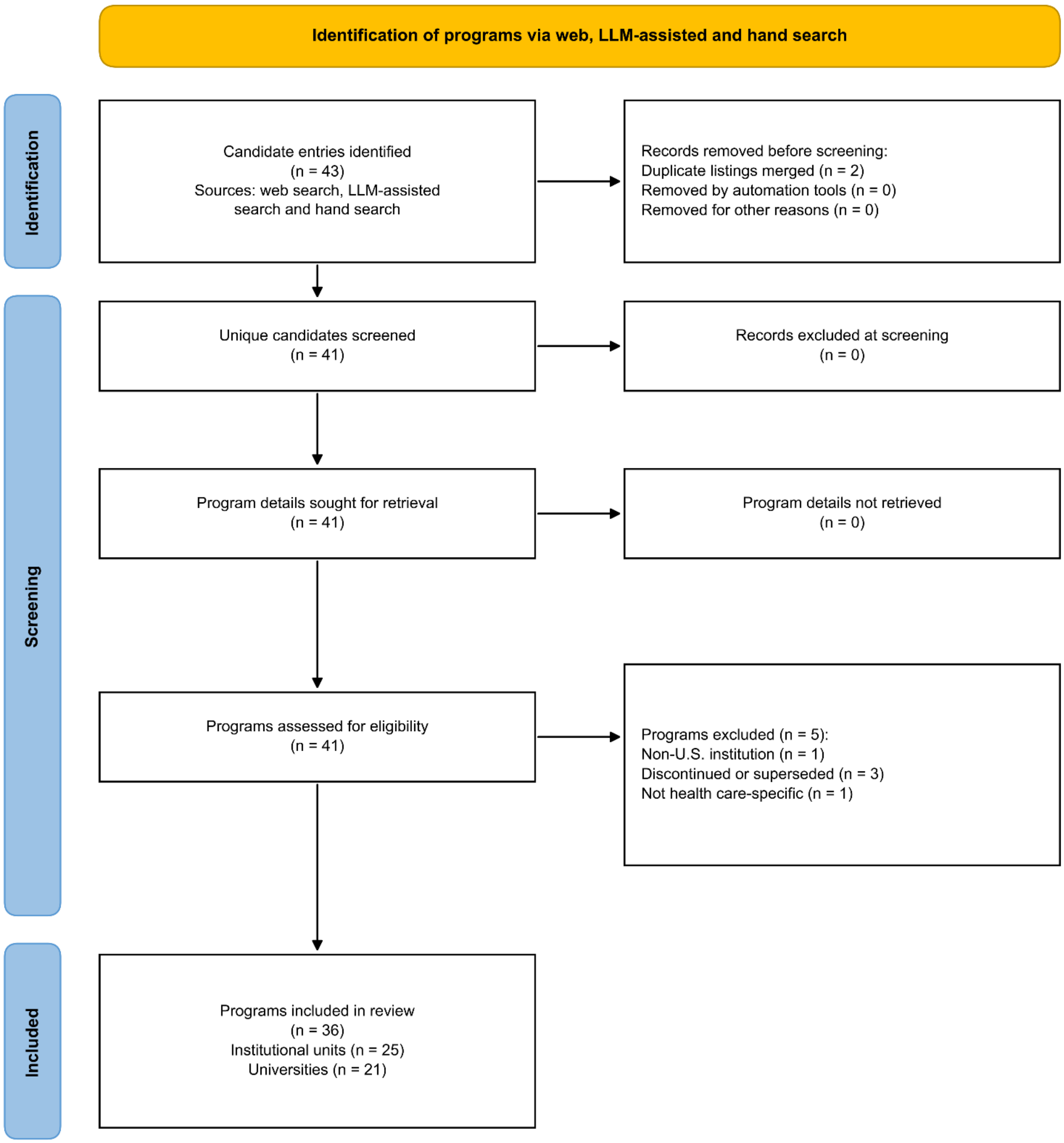

**Fig 1. Identification, screening, and inclusion of programs. Candidate programs were identified through web searches, retrieval-augmented large language model assistants, and hand searches of prior reviews between April and July 2026. Programs run by distinct schools, campuses, or continuing-education arms of one university were counted as separate institutional units. The flow diagram was adapted from the PRISMA 2020 template, licensed under CC BY 4.0. The screening register, including excluded candidates and reasons for exclusion, is provided as S1 Data.**

## Results

### *Programs and Instructors*

The 36 programs included graduate degrees, certificates, executive and professional education, undergraduate courses, and continuing medical education; 26 (72.2%) offered at least partial online access. Most programs described target audiences spanning clinical, technical, or policy backgrounds. Named instructors were identified for 30 programs. Of 92 named instructors, 36 (39.1%) had backgrounds in computer science, machine learning, or engineering, 28 (30.4%) in biomedical informatics or data science, and 6 (6.5%) in statistics, for a combined computational share of 76.1% (Table 1). Clinical medicine accounted for 15 instructors (16.3%), health policy or public health for 4 (4.3%), and the social sciences or humanities for 3 (3.3%). Gender could be determined for 51 instructors, of whom 34 (66.7%) were men; women were most represented in biomedical informatics and data science (8 of 17).

**Table 1. Disciplinary Background and Gender of Named Instructors.**

| Characteristic | Value |
|---|---|
| Named instructors — no. | 92 |
| Primary disciplinary background — no. (%) | |
| Computer science, machine learning, or engineering | 36 (39.1) |
| Biomedical informatics or data science | 28 (30.4) |
| Medicine | 15 (16.3) |
| Statistics or biostatistics | 6 (6.5) |
| Health policy or public health | 4 (4.3) |
| Social sciences or humanities | 3 (3.3) |
| Computational disciplines combined* | 70 (76.1) |
| Gender — no. (%)† | |
| Male | 34 (66.7) |
| Female | 17 (33.3) |
| Could not be determined — no. | 41 |

Instructors named as faculty, lecturers, or course directors on program pages were counted once regardless of the number of programs taught; administrators, coordinators, and teaching assistants were excluded. Named instructors were identified for 30 of the 36 programs. De-identified instructor-level data are provided in Table S4 in S1 Appendix.

* Computer science, machine learning, or engineering; biomedical informatics or data science; and statistics or biostatistics.

† Percentages are among the 51 instructors whose gender could be determined from the pronouns used in public institutional biographies.

### *Collaboration and Admission*

Eighteen programs (50.0%) listed team-based projects, capstones, or mentored research, 9 (25.0%) limited peer interaction to discussion, and 9 (25.0%) reported nothing about collaboration (Table 2). Only 5 programs (13.9%; 5 of the 18 with team-based work) explicitly assembled teams across clinical and technical backgrounds or paired them with clinical mentors. Ten programs (27.8%) required prior technical coursework or proficiency for admission.

**Table 2. Collaborative Learning Structure and Technical Admission Prerequisites (N=36).**

| Characteristic | Programs - no. (%) |
|---|---|
| Collaborative structure | |
| Team-based projects, capstones, or mentored research | 18 (50.0) |
| Teams assembled across clinical and technical backgrounds* | 5 (13.9) |
| Discussion only | 9 (25.0) |
| No group work, explicitly stated | 0 |
| Not reported | 9 (25.0) |
| Technical prerequisites for admission | |
| Coursework or proficiency required | 10 (27.8) |
| Assumed or preferred | 5 (13.9) |
| No technical prerequisite | 9 (25.0) |
| Not reported | 12 (33.3) |

Categories are defined in the Methods section; program-level codes are provided in Table S1 in S1 Appendix.

* Subset of team-based programs (5 of 18 [27.8%]); includes teams paired with clinical mentors.

***Responsible AI Domains***

Ethics was covered by 29 programs (80.6%) (Table 3). Data provenance at the level of the primary outcome was covered by 8 programs (22.2%): 6 (16.7%) at the structural level (the Stanford AI in Healthcare Specialization, two MIT courses [6.7930/HST.956 and HST.953], the Ohio State Graduate Certificate in AI in Digital Health, UC San Diego Principles of Biomedical Informatics, and University of Florida PHC 3621) and 2 (5.6%) at the dataset-limitations level only (Yale Medical Software and Medical AI and University of Florida PHC 4796C). Three further programs (8.3%) taught data management only; with these included, coverage was 11 programs (30.6%) (Table S3). None of the seven master's degree programs and none of the six executive or professional-development programs met the primary outcome. Of the 29 programs covering ethics, 22 did not meet the provenance outcome, and no program addressed sustainability.

**Table 3. Coverage of Responsible AI Domains (N=36).**

| Domain | Programs - no.(%) | 95% CI |
|---|---|---|
| Ethics | 29 (80.6) | 65.0 to 90.2 |
| Data provenance, primary outcome* | 8 (22.2) | 11.7 to 38.1 |
| Structural provenance | 6 (16.7) | 7.9 to 31.9 |
| Dataset limitations only | 2 (5.6) | 1.5 to 18.1 |
| Data management only† | 3 (8.3) | 2.9 to 21.8 |
| Sustainability | 0 | 0.0 to 9.6 |

Confidence intervals (CIs) were calculated with the Wilson score method and have not been adjusted for multiplicity.

* Structural provenance or dataset limitations, as defined in the Methods section.

† Not counted toward the primary outcome. With these programs included, coverage was 11 programs (30.6%; 95% CI, 18.0 to 46.9) (Table S3 in S1 Appendix).

## Discussion

In this analysis of 36 health AI education programs at U.S. universities, ethics was taught in four of every five programs, but how clinical data come to exist was taught in about one in five and, at the structural level, in one in six. Provenance appeared mainly in single courses and in none of the master's degree programs, and no program addressed the environmental cost of computation. We read these findings as a single pattern, in which a field that inherited its educational template from computer science treats health data as a neutral technical input. Within that template, ethics becomes a compliance layer added after the technical core, and the material costs of computation fall outside the discipline.

Regulation, procurement, and postmarket surveillance all act after a system is built. Education comes first, and it shapes which questions a practitioner thinks to ask. The biases described by Obermeyer et al. and Gichoya et al. were found by investigators who examined where their data came from and what they measured [2,3], and practitioners undervalue data work relative to model work partly because of how they are trained [12]. If the field wants practitioners who interrogate data, it will have to teach them to do so; most of the prominent programs we examined do not. Health datasets also carry forward what Fricker calls epistemic injustice, in which testimony is discounted and experiences go missing from the shared record [16], and a curriculum that never examines this trains researchers to inherit it. Making health AI safe and equitable will also take audits, regulation, procurement standards, and community governance. Of these interventions, education has received the least systematic scrutiny.

Three quarters of named instructors came from computational disciplines and 3 of 92 from the social sciences or humanities. More than a quarter of programs required technical preparation for admission, which may screen out clinicians, community health workers, and social scientists. Prerequisites that exclude the people closest to care delivery keep some questions out of the classroom, and a single epistemic tradition comes to be treated as complete [17]. Collaborative pedagogy was common, yet only five programs used it to bring model builders into contact with those who have watched clinical data being made. Such contact matters when it changes which questions students learn to ask and whose knowledge they treat as relevant.

HST.953 has students work alongside intensive care clinicians to understand how the data in a record were generated, and PHC 3621 devotes a semester to whose data are used and how biased data produce discriminatory algorithms. Courses like these show that the gaps we found reflect choices about what to teach. The European Union AI Act now requires providers of general-purpose AI models to document energy consumption [18], yet none of the 36 programs teaches the environmental cost of computation. The Lancet Commission criticized fragmented and static curricula and called for health professional education matched to population needs [13], and Schön described professional expertise as continuing reflection in action [19]. For a program, the equivalent would be to publish complete syllabi with stated outcomes and to revisit content on a fixed cycle. Datasheets established a documentation norm for datasets [20], and a similar norm for curricula would let outside readers check whether a program covers provenance and when its content was last reviewed. Recurring public measurement shapes what a field treats as progress [21], which makes the choice of what to measure a normative one. We intend this survey as the first of an annual series.

The main limitation of this study is that we analyzed advertised curricula, which are an imperfect proxy for classroom practice. Some programs may teach provenance without advertising it. Marketing language tends to overstate coverage of socially valued topics, however, and our ethics coding credited any public mention, so the gaps we report are probably lower bounds. Programs whose documentation was silent on a domain were counted as not covering it, so programs with sparse public materials may be misclassified. Provenance coding is interpretive; all programs were double-coded, but agreement could not be quantified, and coverage was 16.7% under the narrowest definition and 30.6% under the broadest (Table S3). The search was not exhaustive and favored programs with a substantial web presence, and programs launched after July 2026 are not captured. Named instructors were unavailable for six programs. Disciplinary background was taken from public biographies and may not reflect what an instructor teaches. Gender could not be determined for 41 of the 92 instructors, and pronouns in a biography are an imperfect proxy for gender identity.

## Conclusions

Prominent U.S. programs that train the health AI workforce widely taught ethics but rarely taught how clinical data are produced. None taught the environmental cost of computation, and three quarters of named instructors came from computational disciplines.

## Use of artificial intelligence tools

Two retrieval-augmented large language model assistants (Perplexity and Claude, Anthropic) were used during the search to generate candidate programs for screening, as described in Materials and methods; they were not used as a source of evidence about any program, and every candidate was verified by the authors against primary institutional sources. Claude (Anthropic) was also used to edit and condense the manuscript text and to assist with formatting. The authors reviewed every suggestion, verified all data and statements against the underlying dataset, and take full responsibility for the content of this manuscript.

## Supporting information

**S1 Appendix.** Supplementary tables. Table S1, full program dataset (36 programs); Table S2, search queries and prompts; Table S3, sensitivity of the data provenance estimate to the definition applied; Table S4, de-identified instructor-level dataset (92 named instructors).

**S1 Data.** Screening register of all 43 candidate entries, with inclusion decisions and reasons for exclusion.

**S2 File.** Completed PRISMA extension for Scoping Reviews (PRISMA-ScR) checklist.

All supporting information files listed above (S1 Appendix, S1 Data, and S2 File), together with the coded dataset and a codebook describing all coded variables, are openly available on Zenodo at https://doi.org/10.5281/zenodo.22936034.